\documentstyle[12pt]{article}
\newcommand{\Z}{{\sf Z} \! \! {\sf Z}}
\renewcommand{\tiny}{\rm}
\newcommand{\be}{\begin{equation}}
\newcommand{\ee}{\end{equation}}
\newcommand{\bq}{\begin{eqnarray}}
\newcommand{\eq}{\end{eqnarray}}
\newcounter{saveeqn}
\newcounter{App} %\setcounter{App}{0}

\begin{document}

\pagestyle{empty}
\hskip 3.0truein{DFUPG 1/93}
\vskip 0.2truein
\begin{center}
{\large \bf SU($N$) ANTIFERROMAGNETS AND STRONGLY COUPLED QED:
EFFECTIVE FIELD THEORY FOR JOSEPHSON JUNCTIONS ARRAYS}\\
\vspace{1 cm}
{\large M. C. Diamantini}$^1$ {\large and P.
Sodano}\footnote{This work is supported in part by a grant from the
M.U.R.S.T..  M.C.D. and P.S. acknowledge the hospitality of the Physics
Department at the University of British Columbia where some of this
work was completed.}\\
\vspace{0.4 cm}
{Dipartimento di Fisica and Sezione I.N.F.N., Universit\'a di
Perugia\\ Via A. Pascoli, 06100 Perugia, Italy}\\

\vspace{0.5 cm}
{\large E. Langmann}\footnote{Work supported in part by the ``Fonds
zur F\"orderung der wissenschaftlichen Forschung'' of Austria under
contract Nr.  J0789-PHY.} {\large and G. W. Semenoff}\footnote{This
work is supported in part by the Natural Sciences and Engineering
Research Council of Canada.  G.S.  acknowledges the hospitality of the
Physics Department of the University of Perugia and I.N.F.N., Sezione
di Perugia where part of this work was completed.}\\
\vspace{0.4 cm}
{ Department of Physics, University of British Columbia\\
Vancouver, B.C., Canada V6T 1Z1} \\
\vspace{0.7 cm}
{\sl To appear in "Common Trends in Condensed Matter and High Energy
Physics"} \\
{\sl Chia Laguna, Italy, September 1992}
\end{center}
\vskip 0.2truein
\begin{center}
{\bf Abstract}
\end{center}
We review our analysis of the strong coupling limit of
compact QED on a lattice with staggered Fermions.
We show that, for infinite coupling, compact QED is exactly mapped
in a quantum antiferromagnet.
We discuss some aspects of this correspondence relevant for effective field
theories of Josephson junctions arrays.

\vskip 0.3truein
\pagestyle{plain}
\setcounter{page}{1}

\section{Introduction}

\vskip 0.2truein

The strong coupling limit of gauge theories such as QCD exhibits many
of the qualitative features of their spectrum.  For example,
confinement is manifest and also, in some cases, it is straightforward
to show that chiral symmetry is broken.  Recently this issue has been
re-examined in detail \cite{s,dsls} and the possibility for chiral
symmetry breaking for a wide variety of gauge theories has been
explored.

It has been recognized for some time that the strong coupling limit of
lattice gauge theory with dynamical fermions is related to certain
quantum spin systems.  This is particularly true in the Hamiltonian
picture and appeared in some of the earliest analyses of chiral
symmetry breaking in the strong coupling limit \cite {smit}.  In that
work, the chiral symmetry breaking phase of strong coupling gauge
theory was analyzed by drawing an analogy with quantum
antiferromagnets and then using mean field theory and spin-wave
analysis to compute the properties of the pion spectrum.

In more recent literature \cite{fradbk}, it has been noted that
there are several formal similarities between some condensed matter
systems with lattice fermions (particularly certain antiferromagnetic
spin systems) and lattice gauge theory systems, usually in their
strong coupling limit.  For example, it is well known that the quantum
spin 1/2 Heisenberg antiferromagnet is equivalent to the strong
coupling limit of a particular U(1) lattice gauge theory \cite
{Affleck 1988}.  It can also be written as a kind of SU(2) lattice
gauge theory \cite {azha}.

An important feature of the relationship of lattice gauge theory with
spin systems is the fact that the staggered Fermions which are used to
put the Dirac operator on a lattice resemble ordinary lattice fermions
used in tight binding models in condensed matter physics when the
latter have a half-filled band and are placed in a background U(1)
magnetic field $\pi$ (mod $2\pi$) (1/2 of a flux quantum) through
every plaquette of the lattice.  This is an old result for $d=2+1$
\cite {hop} where it was already recognized in the first work on the
Azbel-Wannier-Hofstaeder problem and has since been discussed in the
context of the so-called flux phases of the Hubbard and quantum
Heisenberg models.  It is actually true for all $d\geq 2+1$ and this
equivalence was used in \cite{s,dsls} to find mappings between strong
coupling lattice gauge theories with a wide variety of fermion
contents and various quantum spin systems.

A place in condensed matter where the magnetic flux which is necessary
to produce a relativistic spectrum for the fermions can be achieved is
in the flux phases \cite{fradbk}.  In this case the magnetic field
comes from a certain condensate.

Alternatively, as an external magnetic field, for ordinary lattice
spacings, it is as yet an experimentally inaccessible flux density.
Flux densities of this order of magnitude can presently only be
achieved in analog experiments where macroscopic arrays of Josephson
junctions take the place of atoms at lattice sites.  These systems are
described in mean field theory by tight binding electrons in a
background U(1) magnetic field \cite {ralu,shst} with accuracy
increasing with $d$ \cite {halsey}. For $d=2$ their ground state is
expected to be a flux phase \cite {scbi,diaso}. The low energy limit
in this phase resembles strongly coupled QED.

In the following we shall review some of the features of strongly
coupled QED which are useful in mapping the strong coupling limit
onto spin systems.  We shall also emphasize some of those features
which we feel are valuable in formulating effective field
theories for Josephson junction arrays.  Let us begin with a brief
summary of the physics of the latter arrays.  In the classical limit
they are described by the effective Hamiltonian
\be
H=-J\sum_{<i,j>}\cos(\theta_i-\theta_j+a_{ij})
\ee
where $\theta_i$ are phases of the superconducting order parameter at
the $i$'th lattice site (which is an island of superconducting material
joined by a weak connection to other similar islands), $a_{ij}$ is
proportional to the line integral of the electromagnetic vector
potential along the connection between sites $i$ and $j$,
$$
a_{ij}=\frac{2e}{\hbar c}\int_i^j A\cdot {\rm d}l
$$
Also, $J$ is a coupling constant which is generally a function of
external magnetic field and temperature and the summation is over
nearest neighbors.

The uniformly frustrated system occurs when the external magnetic
field is a constant in each plaquette (where we assume that the
islands lie on a two dimensional lattice),
\be
\sum_{\Box}a_{ij}=2\pi f
\ee
and $f$ is the ratio between the magnetic flux threading the plaquette
and the unit flux.
The latter system can be mapped onto a classical
Coulomb gas problem with Hamiltonian \cite{jknn}
\be
H=-\pi J \sum_{r,r'}\bigl[m(r)+f\bigr]G(r-r')\bigl[m(r')+f\bigr]
\ee
where the charges $m(r)$ reside on the dual lattice, are restricted to
integer values ($m\in\Z$) and the total charge is constrained as
$$
\frac{1}{|V|}\sum_r m(r)=f
$$ ($|V|$ is the total number of lattice sites). In $D=2$ the charges
represent vortices in the phase field of the superconducting order
parameter.  Even in that case, the ground state of this system is
known only for some particular filling fractions $f$ which depend on
the lattice.  For example, on a square lattice the ground states are
known for some rational filling fraction $f=\frac{p}{q}$, $p,q\in\Z$
\cite{halsey,bishop}.  Of particular interest to us is the fully
frustrated case, $f=\frac{1}{2}$ on a square lattice.  There, the
ground state is a staggered configuration with alternating
$m+f=\frac{1}{2}$ and $m+f=-\frac{1}{2}$ on neighboring sites.

Also, for this charged plasma in the continuum, it is known that the
ground state of a classical Coulomb gas of this kind is a triangular
Wigner lattice \cite{halsey}.  Furthermore, when $f=\frac{1}{2}$ there
is a U$(1)$ symmetry and the ground state exhibits a discrete symmetry
$\Z_2$.  At zero temperature the system chooses one of the two phases.

It has also been shown that the equation which determines the
configuration of the order parameter is equivalent to the equation
which determines the spectrum of electrons in a lattice in a
background magnetic field, the so-called Azbel-Wannier-Hofstaeder
problem \cite{ralu,shst}.

These classical systems are the limit of a quantum system where the
variables conjugate to the phases of the order parameters are the
electric charges residing on the islands
\be
\bigl[ \theta_i,Q_j\bigr]=2ie\hbar\delta_{ij}
\ee
The effective Hamiltonian is
\be
H=\frac{1}{2}\sum_{i,j}(Q_i-Q_{xi})C_{ij}^{-1}(Q_j-Q_{xj})
-J\sum_{<i,j>}\cos(\phi_i-\phi_j+a_{ij})
\label{jj}
\ee
where $Q_i$ is the electric charge and $Q_{xi}$ is an offset, or
background charge residing on island $i$. The Coulomb interaction of
the charges is described by the capacitance matrix $C_{ij}$.  The
classical Hamiltonian is regained in the limit
$C_{ij}\rightarrow\infty$ or $J\rightarrow\infty$.

Another model which is often compared with (\ref{jj}), and in fact
which has been argued to describe the critical behavior of non-classical
junction arrays \cite{xxx}, is the
Bose-Hubbard model with Hamiltonian
\be
H=-\frac{J}{2}\sum_{<i,j>}\left(\Phi^{\dagger}_ie^{ia_{ij}}\Phi_j+{\rm
h.c.}\right) +\frac{1}{2}\sum_{ij} V_{ij}n_i(n_j-\delta_{ij})
-\mu\sum_i n_i
\ee
where $\Phi_i$ and $\Phi_i^{\dagger}$ are the annihilation and
creation operator for a boson at site $i$, respectively and
$n_i=\Phi^{\dagger}_i\Phi_i $ is the boson number at site $i$.
The total number of bosons is controlled by the chemical potential
$\mu\leq 0$.
The on-site part of the self-interaction vanishes if the site is
occupied by a single boson.   The energy for more than one
boson on a site should be large, to give the bosons a hard core.
With hard core bosons, this model is equivalent
to the frustrated quantum XY model \cite{diaso}.

There is also  a mapping of the frustrated quantum XY model onto a
lattice Chern-Simons theory coupled to Fermions \cite{diaso}.
For the uniformly frustrated case, the mean field theory of the
latter system
is equivalent to lattice fermions in an external field, related
to the frustration parameter.  The fermion
spectrum is that of the Hofstaeder problem and  is similar to
the fermion system
encountered in the
flux phase of the Hubbard model.

It is an interesting question to what extent these models share other
features of a lattice field theory such as quantum electrodynamics.
If the fermion spectrum is relativistic and, given that the models
have a U(1) gauge invariance, one might expect dynamical effects to
generate the relevant operators of electrodynamics.
Furthermore, the infinite coupling limit of lattice QED reduces to the
problem of solving for the ground state of a classical coulomb gas with
a logarithmic potential, identical to the classical limit of the junction
array problem.  The ground state of this coulomb gas system has been
examined in both contexts in previous literature.

It is interesting to speculate that, when we move away from the
classical limit in the case of junction arrays, and when we move away
from the infinite coupling limit in the case of compact QED, we
still have systems with many common features.  For example, both
systems have a first order phase transition.  In the case of the
juction array it is the transition which destroys superconductivity.
In the case of compact electrodynamics it is where chiral
symmetry\footnote{We shall see in following sections that, for
staggered fermions, chiral symmetry corresponds to invariance under
translations by one lattice site}, which is broken in the strong
coupling limit, is restored. It is accompanied by a
confinement-deconfinement transition for electric charges.  In the
coulomb gas representation of the classical junction array the phase
transition corresponds to melting of the Wigner lattice to form a two
dimensional plasma.  In compact electrodynamics the phase transition
restores full translation invariance of the lattice theory
(corresponding to chiral invariance in the continuum theory) and the
electrons are deconfined. In the latter case the strong coupling,
confining ground state has the structure of a Wigner lattice of
electrons.

In the following we shall review some aspects of strongly coupled
lattice QED relevant to an effective field theory of Josephson
junction arrays. We shall show how the analogy between Dirac
fermions and
tight-binding condensed matter fermions is used to find an exact
mapping of the strong coupling limit of compact lattice QED with
dynamical staggered fermions onto certain quantum SU$(N)$
antiferromagnets.  The precise structure of the resulting
antiferromagnet depends on the number of fermion flavors.  We shall also
review the argument \cite{s,dsls} by which QED with staggered Fermions
in any number of
spatial dimensions, $D>1$, resembles a condensed matter system of
lattice electrons in an external magnetic field and with a half-filled
band .

Of course, the understanding of the phase structure, particularly the
confining and chiral symmetry breaking phases, and nature of phase
transitions of QED is of great fundamental interest to particle
physicists. These issues have been studied in the continuum quantum
electrodynamics where the possibility of a phase transition in the
strong coupling regime is suggested by approximate analytical results
using the quenched ladder approximation.  The idea for the mechanism,
which has long been advocated by Miranski and collaborators\cite{fomin
et.al.,miransky}, is that when the electric charge reaches a certain
critical value the vacuum of QED becomes unstable, leading to a sort
of collapse phenomenon, analogous to the sparking of the vacuum in the
presence of a supercritical nucleus.  In the picture which we propose,
the analog of the collapse phenomenon is the formation of either
charge or isospin density waves and the resulting reduction of the
lattice translation symmetry from translations by one site to
translations by two sites.  It occurs when the exchange interaction of
the electrons, which is attractive, dominates the tendency of the
direct Coulomb interaction and the kinetic energy to delocalize
charge, giving an instability to the formation of commensurate charge
density waves.  This forms a gap in the Fermion spectrum and a
particular mass operator obtains a vacuum expectation value.  This
gives an intuitive picture of how strong attractive interactions in a
field theory can form a coherent structure.  Here, the commensurate
density waves in the condensed matter system correspond to a
modulation of the vacuum charge or isospin density at the ultraviolet
cutoff wavelength in the field theory.

In the continuum, we are generally interested in massless quantum
electrodynamics with the action
\bq
S=\int d^{D+1}x\left(-\frac{\Lambda^{3-D}}{4e^2}F_{\mu\nu}F^{\mu\nu}
+\sum_{a=1}^{N_F}\bar\psi^a\gamma^\mu(i\partial_\mu+A_\mu)\psi^a\right)
\label{contqed}
\eq
where $\Lambda$ is the ultraviolet cutoff, $e$ is the dimensionless
electric charge and there are $N_F$ flavors of
$2^{[(D+1)/2]}$--dimensional Dirac spinors. (Here, $[(D+1)/2]$ is the
largest integer less than or equal to $(D+1)/2$.)  In even dimensions
(when D+1 is even), the flavor symmetry is
SU$_R(N_F)\times$SU$_L(N_F)$.  In odd dimensions there is no chirality
and the flavor symmetry is SU$(N_F)$.  What is usually referred to as
chiral symmetry there is a combination of parity and a $\Z_2$ subgroup
of the flavor group.  We shall use a lattice regularization of
(\ref{contqed}) and study the limit as $e^2\rightarrow\infty$.  We
shall use $N_L$ flavors of staggered fermions.  In the naive continuum
limit, this gives $N_F=N_L~$ 2--component fermions in $D=1$, $N_F=2N_L~$
2--component fermions in $D=2$ and $N_F= 2N_L~$ 4--component fermions in
$D=3$.  Though the lattice theory reduces to (\ref{contqed}) in the
naive continuum limit, the lattice regularization breaks part of the
flavor symmetry.  All of the operators which are not symmetric are
irrelevant and vanish as the lattice spacing is taken to zero.  In
$D=3$, the SU$_R(N_F)\times$SU$_L(N_F)$ symmetry is reduced to
SU$(N_F/2)$ (in $D=3$, $N_L=N_F/2$) and translation by one site in the 3
lattice directions.  These discrete transformations correspond to
discrete chiral transformations in the continuum theory.  In $D=2$, the
lattice has SU$(N_F/2)$ (in $D=2$, $N_L=N_F/2$) symmetry and two
discrete (``chiral'') symmetries.  In $D=1$, the
SU$_L(N_F)\times$SU$_R(N_F)$ symmetry is reduced to SU$(N_F)$ (in $D=1$,
$N_L=N_F$) and a discrete chiral transformation.  In each case, the
discrete chiral symmetry is enough to forbid Fermion mass and it is
the spontaneous breaking of this symmetry which we shall examine and
which we shall call ``chiral symmetry breaking''.  We shall also
consider the possibility of spontaneous breaking of the SU$(N_F)$
flavor symmetry.  In the continuum the SU$(N_F)$ corresponds to a
vector--like symmetry subgroup of the full flavor group.  Since the
lattice and continuum theories have different symmetries, the spectrum
of Goldstone bosons, etc. would be different in the two cases. We
shall not address the source of these differences but will define QED
by its lattice regularization and discuss the realization of the
symmetries of that theory only.

In the strong coupling limit,
lattice QED with $N_{\tiny L}$ lattice flavors of staggered
Fermions is {\bf
exactly equivalent} to an SU$(N_{\tiny L})$ quantum antiferromagnet
where the spins are in a particular fundemental representation of the
SU$(N_{\tiny L})$ Lie algebra. Furthermore, mass operators of QED
correspond to staggered charge and isospin density operators in the
antiferromagnet.  Thus, the formation of charge-density waves
corresponds to chiral symmetry breaking and the
dynamical generation of an iso--scalar Fermion mass whereas
N\'eel order of the antiferromagnet
corresponds to dynamical generation of Fermion mass with an
iso--vector condensate.    As
a result of this correspondence
we are able to use some of the properties of the quantum
antiferromagnets to deduce features of strongly coupled QED.

In the infinite coupling limit, the properties of the
electronic ground state of compact and non-compact QED are identical.
The fact that compact QED confines and non-compact QED (at least in high
enough dimensions) does not confine affects only the properties of the gauge
field wavefunctions and the elementary excitations. In the following we
shall focus our attention only on compact QED.
\vskip 0.1truein

\noindent
An interesting result of our approach \cite {dsls} is that the case
of even $N_{\tiny L}$ and odd $N_{\tiny L}$ are very different:
\vskip 0.1truein

When $N_{\tiny L}$ is odd, the vacuum energy in the strong coupling
limit is proprotional to $e^2$, the square of the electromagnetic
coupling constant.  We also find that chiral symmetry is broken in the
strong coupling limit for all odd $N_{\tiny L}\geq 1$ and for all
spacetime dimensions $D+1\geq2$.  The mass operator which obtains a
nonzero expectation value is a Lorentz and flavor Lie algebra scalar.
There are also mass operators which are Lorentz scalars and which
transform nontrivially under the flavor group which could get an
expectation value and break the flavor symmetry spontaneously if
$N_{\tiny L}$ is small enough. When
$N_{\tiny L}$ increases to some critical value there is a phase
transition to a disordered phase \cite {diasod}.

In contrast, when $N_{\tiny L}$ is even we find that the vacuum energy
in the strong coupling limit is of order 1.  We find that chiral
symmetry is broken when the spacetime dimension
$D+1\geq 3$ and when $N_{\tiny L}$ is small enough.  The mass operator
which gets an expectation value is a Lorentz scalar and transforms
nontrivially under the flavor SU$(N_{\tiny L})$.  Thus, flavor
symmetry is spontaneously broken.  As in the case of odd $N_{\tiny
L}$, there is an upper critical $N_{\tiny L}$ where
there is a transition to a disordered phase \cite {diasod}.

For $N_{\tiny L}$ odd finding the gauge
invariant electronic ground state of strongly coupled lattice QED is
equivalent \cite {dsls} to finding the exact ground state of the
generalized classical Coulomb gas model in $D$--dimensions with Hamiltonian
\bq
H_{\rm coul}=\frac{e^2}{2}\sum_{<x,y>}\rho(x)g(x-y)\rho(y)
\eq
where the variable $\rho(x)$ lives at the sites of a cubic lattice with
spacing one (with coordinates $(x_1,\ldots,x_D)$ and $x_i$ are integers)
and takes on values
\bq
-\frac{N_{\tiny L}}{2},-\frac{N_{\tiny L}}{2}+1,\ldots,\frac{N_{\tiny L}}{2}+1,
\frac{N_{\tiny L}}{2}
{}~~N_{\tiny L}~{\rm an~odd~integer}
\eq
and the interaction is long-ranged
\bq
g(x-y)\sim\vert x-y\vert^{2-D}~~,~~{\rm as}~\vert x-y\vert\rightarrow\infty
\eq
In any space dimension, $D\geq1$ we prove that there are two degenerate
ground states which have the
Wigner lattice configurations
\bq
\rho(x)=\pm {1\over2}(-1)^{\sum_{i=1}^D x_i}
\eq
When $N_{\tiny L}=1$ this is a long-ranged Ising model which is known
to have antiferromagnetic order, even in one dimension.  It is also in
agreement with a known result about the Ising model in two dimensions
\cite{halsey}.

The Coulomb gas representation of the classical Josephson junction
array is related to this system when $f=1/2$, i.e. in the fully
frustrated case.  There, the charge on a site is only limited by
dynamics, whereas in electrodynamics it is restricted by the Pauli
principle to be between $-N_L/2$ and $N_L/2$.  However, in the case of
the classical Coulomb gas, the Coulomb energy of more than one charge
residing on a lattice site is infinite, so effectively, whenever $f<1$
the charges at lattice sites should also be either $1,0$ or $-1$.
Therefore the problem of determining the charge (vortex) distribution
in a fully frustrated junction array is equivalent to finding the
ground state of infinite coupling QED with $N_L=1$.  It will turn out
that the solution of this problem is similar for all odd $N_L$.

The difference between even and odd $N_{\tiny L}$ can be seen to arise
from a certain {\it frustration} encountered when\footnote{Note that
this {\it frustration} is different from that caused by the external
magnetic field in the junction array}, with odd $N_{\tiny L}$, one
simultaneously imposes the conditions on lattice QED which should lead
to charge conjugation invariance, translation invariance and Lorentz
invariance of the continuum limit.  This {\it frustration} is absent
when $N_{\tiny L}$ is even.  This is not an anomaly in the
conventional sense of the axial anomaly or a discrete anomaly
encountered in the quantization of gauge theories, as no symmetries
are incompatible, but it is nevertheless an interesting analog of the
anomaly phenomenon.

Note that in 2+1 dimensions the difference between even and odd
$N_{\tiny L}$ is the difference between an even and odd number of
4-component Fermions.  Our result that for an odd number of
4-component Fermions the chiral symmetry is always broken for large
coupling seems to contradict the continuum analysis in
\cite{Applequist,Nash}.  We do not fully understand the reason for
this, but speculate that it is related to the lattice regularization.
Their model is very similar to ours in that they effectively work in
the strong coupling limit when they replace the ultraviolet
regularization which comes from having a Maxwell term in the QED
action by a large momentum cutoff.  In our case we have a lattice
cut-off and the result should be very similar.  Note that we agree
with them when $N_{\tiny L}$ is even and there is an even number of
4-component Fermions.  In that case, we are also in qualitative
agreement with recent numerical work \cite{kogutnew} which, since it
uses Euclidean staggered Fermions and there is a further Fermion
doubling due to discretization of time, can only consider the case
where there is an even number of 4-component Fermions. The anomalous
behavior that we find with odd numbers of 4-component Fermions exposes
a difficulty with treating the flavor number $N$ as a continuous
parameter.

A hint as to why 2+1--dimensional Fermions should come in
four-component units appears when we formulate compact QED in the
continuum using the SO(3) Georgi-Glashow model with spontaneous
breaking of the global symmetry, SO(3)$\rightarrow$U(1) and the limit
of large Higgs mass \cite{poly}.  We begin with the model
\bq
{\cal L}= \frac{1}{4e^2}\sum_{a=1}^3(F^a_{\mu\nu})^2+\frac{1}{2}\sum_{a=1}^3
(D\phi)^a\cdot(D\phi)^a+\frac{\lambda}{4}
((\sum_{a=1}^3\phi^a\phi^a)-v^2)^2
\eq
The spectrum contains a massless photon which in the $\lambda\rightarrow
\infty$ and $v^2\rightarrow\infty$ is the only light excitation.
Since the U(1) gauge group is a subgroup of SO(3), it is compact.  We
wish to make 2+1-dimensional electrodynamics by coupling this model to
Fermions.  It is known that, if the resulting theory is to preserve
parity and gauge invariance simultaneously, we must use an even number
of two-component Fermions which, in the minimal case are also SU(2)
doublets \cite{niemi}
\bq
{\cal L}_F=\bar\psi\left(i\gamma\cdot\nabla+\gamma\cdot A+
\gamma^5(m+g\phi\cdot\sigma)\right)\psi
\eq
Here, the Fermions are four-component and have a parity invariant mass
and Higgs coupling.  Each SU(2) doublet contains two electrons (which
can be defined so that both components of the doublet have the same
sign of electric charge), thus the basic Fermion for compact QED has
eight components.  The maximal chiral symmetry is obtained in the
massless case with vanishing Higgs coupling.  It is possible, by
suitable choice of mass and Higgs coupling i.e. $m=\pm g<\phi>$, (and
also reduction of the chiral symmetry) to make four of the Fermions
heavy, leaving four massless components.  This is consistent with
parity and gauge invariance.  It is also interesting to note that
staggered Fermions on a Euclidean lattice, where time is also
discretized, produce eight component continuum Fermions.

In Section $2$ we review the essential features of the staggered
fermion formalism. In Section $3$ we elucidate how spin algebras
may be represented by Fermion bilinears.  In Section 4 we formulate
lattice QED and analyze its discrete symmetries. Section $5$ is
devoted to the analysis of the strong coupling limit of compact QED.
There we shall review the properties of the electronic ground state
for both $N_{\tiny L}$ even and odd.  Section $6$ is devoted to
concluding remarks.
\section{\bf  Staggered Fermions}

Here we make explicit the fact that latticized (staggered)
relativistic fermions are equivalent to the Hofstaeder problem
with flux density 1/2 flux quanta per plaquette.

In the following, we use units in which the lattice spacing, the speed of
light and Planck's constant are all equal to one.  Space is a
$D$-dimensionsal cubic lattice with lattice points
$x=(x_1,x_2,\ldots,x_D)$ ($x_i$ integers), oriented links
$[x,i]$ labeled by their site of origin and their direction,
$i=\pm1,\dots,\pm D$, and oriented plaquettes $[x,i,j]$ with
corners $x,x+\hat i,x+\hat i+\hat j,x+\hat j$ and orientation $\hat
i\times\hat j$ where $\hat i$ is the unit vector in $i$-direction.

In relativistic field theory, the purpose of the staggered Fermion method
is to minimize Fermion doubling which always occurs when one tries to
put Fermions on the lattice. Generally, staggered Fermions are
obtained by the spin-diagonalization method. To implement this
method, we begin with the naively latticized Dirac Hamiltonian
for free relativistic fermions,
\bq
H_F={1\over2}\sum_{[x,j]}\left(
\psi^{\dagger}(x)i\alpha^j\nabla_j\psi(x)-
(\nabla_j\psi^{\dagger})(x)i\alpha^j\psi(x)\right)
\label{2.1}
\eq
where the lattice derivative operator is defined as
difference operator
\[
\nabla_i f(x)= f(x+\hat i)-f(x)
\]
and $\alpha^j$ are the $2^{[(D+1)/2]}$-dimensional Dirac
$\alpha$-matrices.  (Here $[(D+1)/2]$ is the integer part of
$(D+1)/2$.)  The latter are Hermitean, $\alpha^{j\dagger}=\alpha^j$,
and obey the Clifford algebra
\bq
\left\{ \alpha^i,\alpha^j\right\}=2\delta^{ij}
\label{2.3}
\eq
They are therefore unitary matrices, $\alpha^{i\dagger}\alpha^i=1$.
Using the properties of the difference operator, (\ref{2.1}) can be
presented in the form
\bq
H_F=-{i\over2}\sum_{[x,j]}\left( \psi^{\dagger}(x+\hat j)\alpha^j\psi(x)-
\psi^{\dagger}(x)\alpha^j\psi(x+\hat j)\right)
\label{2.2}
\eq
Since the Dirac matrices are unitary, the naive lattice Fermion
Hamiltonian in (\ref{2.2}) resembles a condensed matter Fermion
hopping problem with a background U$(2^{[(D+1)/2]})$ gauge field given by
the $\alpha$-matrices.  In any plaquette of the lattice, $[x,i,j]$,
this background field has curvature
\bq
\alpha^i\alpha^j\alpha^{i\dagger}\alpha^{j\dagger}=-1
\label{2.4}
\eq
The curvature resides in the U(1) subgroup of U$(2^{[(D+1)/2]})$ and has
exactly half of a U(1) flux quantum per plaquette.  This is true in
any dimensions. We observe that either 1/2 or zero flux quanta are
the only ones allowed by translation invariance and parity and time
reversal symmetries of the Hamiltonian.

Since the curvature of the $\alpha$-matrices is U(1)-valued, we should
be able to do a gauge transformation which presents the matrices themselves
as U(1) valued gauge fields (i.e. diagonal).  A specific example of
such a gauge transformation due to Kluberg-Stern et.al. \cite{Kluberg} is
\bq
\psi(x)\rightarrow (\alpha^1)^{x_1}(\alpha^2)^{x_2}
\dots(\alpha^{D})^{x_{D}}\psi(x)
\label{2.5}
\eq
Then
\bq
\psi^{\dagger}(x+\hat j)\alpha^j\psi(x)\rightarrow (-1)^{\sum_{k=1}^{j-1}x_k}
\psi^{\dagger}(x+\hat j)\psi(x)
\label{2.6}
\eq
The resulting Hamiltonian is
\bq
H_F=-{i\over2}\sum_{[x,j]}(-1)^{\sum_{k=1}^{j-1}x_k}
\left(\psi^{\dagger}(x+\hat j)
\psi(x)-\psi^{\dagger}(x)\psi(x+\hat j)\right)
\label{2.7}
\eq
This describes $2^{[(D+1)/2]}$ identical copies of Fermions with the
same Hamiltonian which must all give Fermions with the same spectrum
as the original Hamiltonian in (\ref{2.1}).  Staggered Fermions are
obtained by choosing one of these copies.  This reduces the Fermion
doubling by a factor of the dimension of the Dirac matrices,
$2^{[(D+1)/2]}$.

In the staggered Fermion method, we treat the components of the
original lattice Dirac Hamiltonian as flavors, rather than components
of the relativistic spinor necessary for Lorentz invariance.  The
spinor components now reside on adjacent lattice sites.  In this
method, the continuous chiral symmetry of the massless Hamiltonian,
under the transformation $\psi\rightarrow e^{i\gamma^5\theta}\psi$, is
lost. There is a discrete chiral symmetry, corresponding to
translations by one lattice site in any direction.  Explicitly,
\bq
\psi(x)\rightarrow (-1)^{\sum_{k=j+1}^{D}x_k}\psi(x+\hat j)
\label{2.8}
\eq
is a symmetry of the Hamiltonian (\ref{2.7}) and corresponds to a
discrete chiral transformation.

Mass operators correspond to staggered charge densities.  The operator
\bq
\Sigma = \sum_x (-1)^{\sum_{k=1}^{D}x_k}\psi^{\dagger}(x)\psi(x)
\label{2.9}
\eq
changes sign under the chiral transformations (\ref{2.8}) and
corresponds to a certain Dirac mass.

With staggered Fermions there is still a certain amount of Fermion
doubling.  The doubling can be counted by noting that the staggered
Fermion Hamiltonian (\ref{2.7}) is invariant under translations by two
lattice sites.  Therefore, a unit cell is a unit hypercube of the
lattice, containing $2^{D}$ sites and staggered Fermions correspond
to a $2^{D}$ component spinor.  The dimension of the Dirac matrices
is $d^{[(D+1)/2]}$.  Therefore the number of Dirac spinors we obtain is
$2^{D}/2^{[(D+1)/2]}$.  For lower dimensions the minimum number of
continuum flavors can be tabulated as
$$
\matrix{ d&~{\rm dim.~of~Dirac~matrices}~&~{\rm No.~of~flavors}\cr
         1+1&   2& 1\cr
         2+1&   2& 2\cr
         3+1&   4& 2\cr  }
$$
Only in 1+1-dimensions do we get a single species of Dirac Fermion.

To see how to take the continuum limit explicitly, consider the case
of $d=3+1$.  There, we divide the lattice into eight sublattices and
label the spinor components as
\bq
\psi({\rm even},{\rm even},{\rm even})\equiv \psi_1
{}~~~\psi({\rm odd},{\rm even},{\rm odd})\equiv \psi_7
\label{2.11a}
\eq
\bq
\psi({\rm even},{\rm odd},{\rm even})\equiv \psi_6
{}~~~\psi({\rm even},{\rm even},{\rm odd})\equiv \psi_5
\label{2.11b}
\eq
\bq
\psi({\rm odd},{\rm odd},{\rm even})\equiv \psi_4
{}~~~\psi({\rm odd},{\rm even},{\rm odd})\equiv \psi_3
\label{2.11c}
\eq
\bq
\psi({\rm even},{\rm odd},{\rm odd})\equiv \psi_2
{}~~~\psi({\rm odd},{\rm odd},{\rm odd})\equiv \psi_8
\label{2.11d}
\eq
In terms of these spinors, the Hamiltonian (\ref{2.7}) can be written
as the matrix operator
\bq
H=\int_{\tilde\Omega_B}{d^{3}k}
\psi^{\dagger}(k)A^i\sin k_i\psi(k)
\label{2.12}
\eq
where
\bq
\tilde \Omega_B=\{ k_i: -\pi/2<k_i\leq\pi/2\}
\eq
is the Brillouin zone of the (even,even,even) sublattice,
\bq
A^i=\left(\matrix{ 0&\alpha^i\cr\alpha^i&0\cr}\right)
\label{2.13}
\eq
and
\bq
\alpha^1=\left(\matrix{ 0&1\cr 1&0\cr}\right)
{}~~~
\alpha^2=\left(\matrix{ \sigma^1&0\cr 0&-\sigma^1\cr}\right)
{}~~~
\alpha^3=\left(\matrix{ \sigma^3&0\cr 0&-\sigma^3\cr}\right)
\label{2.14}
\eq
are a particular representation of the Dirac matrices.

In this representation the mass operator is
\bq
\sum_x (-1)^{\sum_{k=1}^{3}x_k}\psi^{\dagger}(x)\psi(x)=\int_{\tilde\Omega_B}
\frac{d^{3}k}{(2\pi)^3} \psi^{\dagger}(k)B\psi(k)
\label{2.15}
\eq
where
\bq
B=\left(\matrix{ 1&0\cr 0&-1\cr}\right)
\label{2.16}
\eq

The Fermion spectrum is
\bq
\omega(k)=\sqrt{\sum_{i=1}^3\sin^2k_i+m^2}
\label{fermspec}
\eq
and only the region $k_i\sim0$ is relevant to the continuum limit. We
have normalized $\psi(k)$ so that
\bq
\left\{\psi(x),\psi^{\dagger}(y)\right\}=\delta(x-y)
{}~~,~~
\left\{\psi(k),\psi^{\dagger}(l)\right\}=\delta(k-l)
\label{mscom}
\eq
If we define
\bq
\beta=\left(\matrix{\sigma^2&0\cr0&-\sigma^2}\right)
\eq
and the unitary matrix
\bq
M={1\over2}\left(\matrix{1-\beta&1+\beta\cr1+\beta&1-\beta\cr}\right)
\eq
and
\bq
\psi=M\psi'
\eq with
\bq
\psi'=(\psi_a,\psi_b)
\eq
the Hamiltonian is
\bq
H_f=\int_{\Omega_B}{d^{3}k}
\left(\psi_a^{\dagger},\psi_b^{\dagger}\right)
\left(\matrix{\alpha^i\sin k_i-\beta m&0\cr0&\alpha^i\sin k_i+\beta
m\cr}\right)\left(\matrix{\psi_a\cr\psi_b\cr}\right)
\eq
In the low momentum limit, $\sin k_i\sim k_i$, with Fermion density
1/2 per site so that the Fermi level is at the intersection point of
the positive and negative energy bands, we obtain 2 continuum Dirac
Fermions.

This describes two flavors of 4-component Dirac Fermions and the Dirac
masses for each component given by the staggered charge density have
opposite signs.  Thus the charge density breaks the discrete chiral
symmetry.  It also breaks a flavor symmetry which, in the absence of
mass, mixes the two continuum Fermions. The continuum limit for arbitrary $d$
is discussed in \cite {dsls}.

\section{\bf Fermion Representation of SU($N$) Quantum Antiferromagnet}

In order to set the formalism which allows us to find the exact
mapping between strongly coupled lattice QED and SU($N$) quantum
antiferromagnets, we review here the representation of an
SU($N$) antiferromagnet in terms of fermion bilinears.  The analysis
holds for any space dimension D.

The Hamiltonian for an U($N$) quantum antiferromagnet is
\bq
H_{\rm AFM}=\frac{g^2}{2}\sum_{<x,y>} J_{ab}(x)J_{ba}(y)
\label{AFM}
\eq
where $J_{ab}(x)$, $a,b=1,\cdots,N_{\tiny L}$, obey current algebra
relations associated with  the Lie algebra of U($N$),
\bq
\bigl[ J_{ab}(x),J_{cd}(y)\bigr]=\left(J_{ad}(x)\delta_{bc} -
J_{cb}(x)\delta_{ad}\right)\delta(x-y)
\label{algebra}
\eq
and where $<x,y>$ denotes the link connecting sites $x$ and $y$ on a
bipartite lattice.  For simplicity, we take the lattice to be
cubic.  Here, we have used a particular basis for the SU($N$) algebra
which can be conveniently represented by Fermion bilinear operators,
\bq
J_{ab}(x)=\psi^{a\dagger}(x)\psi^b(x)-\delta^{ab}/2
\label{generators}
\eq
with Fermion field operators $\psi^{a(\dagger)}$ obeying
\[
\left\{\psi^a(x),\psi^{b\dagger}(y)\right\} = \delta_{ab}\delta(x-y)
\]
The representation of the algebra on each site $x$ is fixed by
specifying the Fermion number of the states,
\bq
\rho(x)=\sum_a J_{aa}(x)
\eq
For example, the Fermion
vacuum state $\vert0>$ such that
\bq
\psi^a(x)\vert 0>=0~~~,~\forall a,x
\label{vac}
\eq
is the singlet state, the states with $m\leq N$ Fermions per site,
\bq
\prod_x \psi^{a_1\dagger}(x)\psi^{a_2\dagger}(x)\ldots\psi^{a_m\dagger}(x)
\vert 0>
\nonumber
\eq
corresponds to $\rho(x)=m-N/2$ for all $x$ and the irreducible
representation with the Young Tableau with one column and $m$ boxes.

For each site $x$, $\rho(x)$ is the generator of the U(1) subgroup of
U($N$).  Using a basis $T^i=(T^i)^*$, $i=1,\ldots, N^2-1$,
of the Lie algebra of SU($N$) in the fundamental
representation normalized so that ${\rm tr}(T^iT^j)= T^i_{ab}T^j_{ba}
= \delta^{ij}/2$, and using
\bq
T^i_{ab}T^i_{cd}=
\frac{1}{2}\delta_{ad}\delta_{bc}-\frac{1}{2N}\delta_{ab}\delta_{cd}
\label{relSUN}
\eq
it is convenient to introduce
\bq
J^i(x) = \psi^{a\dagger}(x)T^i_{ab}\psi^a(x)
\eq
obeying current algebra of the Lie algebra of SU($N$), and to write
the Hamiltonian (\ref{AFM}) as
\bq
H_{\tiny AFM} = \frac{g^2}{N} \sum_{<x,y>}\rho(x)\rho(y) + H_{{\rm
SU}(N)}
\eq
with
\bq
H_{{\rm SU}(N)} = g^2\sum_{<x,y>} J^i(x)J^i(y)
\eq
is the Hamiltonian of an SU($N$) antiferromagnet.  From this it is
obvious that by fixing the $\rho(x)$, $H_{\tiny AFM}$ is reduced to an
SU($N$) antiferromagnet.

For example, the familiar $j=1/2$ SU(2) Heisenberg antiferromagnet is
obtained from (\ref{AFM}), $N=2$, by using the identity
\bq
\frac{\vec\sigma_{ab}}{2}\cdot\frac{\vec\sigma_{cd}}{2}=\frac{1}{ 2}
\delta_{ad}\delta_{bc}
-\frac{1}{4}\delta_{ab}\delta_{cd}
\label{identity}
\eq
corresponding to (\ref{relSUN}) for $N=2$.

Generally, when $N$ is even we will consider the representations where
$m=N/2$, so that the Fermion occupation of each site is $N/2$ and
$\rho(x)=0$.  When $N$ is odd we divide the lattice into two
sublattices such that the nearest neighbors of all sites of one
sublattice are in the other sublattice (when this is possible the
lattice is said to be bipartite).  When $N$ is odd, the representation
of SU($N$) has $(N+1)/2$ Fermions, i.e. $\rho(x)=1/2$, on the sites of
one sublattice and $(N-1)/2$ Fermions, i.e. $\rho(x)=-1/2$ on the
sites of the other sublattice.

\section{\bf QED on a Lattice}

In this Section we shall review the Hamiltonian formalism of Abelian
gauge fields on a lattice.  For the most part, this formalism can be
found in some of the classic reviews of lattice gauge theory,
\cite{Kogutreview} for example.  A novel feature of our approach \cite
{dsls} is a careful treatment of the normal ordering of the charge
operator and a discussion of the ensuing discrete symmetries.  This
normal ordering turns out to be important if the continuum limit has
to have the correct behavior under $C$, $P$ and $T$ transformations.
It will also be important in our later solution of the strong coupling
limit.

\subsection{\bf Hamiltonian and Gauge Constraint}

We shall discretize space as a cubic lattice and, in order to use the
Hamiltonian formalism; time is left continuous.  Lattice gauge fields
are introduced through the link operators
\bq
U_i(x)\equiv e^{iA_i(x)}
\label{3.1}
\eq
which correspond to the links $[x,i]$ of the lattice.  Electric fields
propagate on links of the lattice and the electric field operator
$E_i(x)$ associated with the link $[x,i]$ is the canonical conjugate
of the gauge field
\bq
\bigl[ A_i(x), E_j(y)\bigr]=i\delta_{ij}\delta(x-y)
\label{3.2}
\eq
The gauge field and electric field operators obey the relations
\bq
A_{-i}(x) = -A_i(x-\hat i)~~~,~~E_{-i}(x)=-E_i(x-\hat i)
\eq
The Hamiltonian of compact QED is
\bq
H_{\rm
NC}=\sum_{[x,i]}{e^2\over2}E^2_i(x)+\sum_{[x,i,j]}{1\over2e^2}\sin^2\left
(B[x,i,j]/2\right) ~~~~~~~~~~~~~~~~~~~~\nonumber\\ +\sum_{[x,i]}\left(
t_{[x,i]}\psi^{a\dagger}(x+\hat i)e^{iA_i(x)}\psi^a(x)+{\rm h.c.}\right)
\label{3.3}
\eq
where the second term contains a sum over plaquettes and the magnetic
field is defined as the curvature of the gauge field at the plaquette
$[x,i,j]$,
\bq
B[x,i,j]=A_i(x)+A_j(x+\hat i)+A_{-i}(x+\hat i+\hat j)+A_{-j}(x+\hat
j)\nonumber\\ =A_i(x)-A_j(x)+A_j(x+\hat i)-A_i(x+\hat j)
\label{3.4}
\eq
As discussed in Section $2$, the hopping parameters $t_{[x,i]}$ contains
phases which produce a background magnetic flux $\pi$ per plaquette.
In the weak coupling continuum limit, the magnitude of $t_{[x,i]}$ is
one, $\vert t\vert^2\equiv\vert t_{[x,i]}\vert^2=1$ in order that the
speed of the free photon and free electron fields are equal, i.e. so
that the low frequency dispersion relations for both the photon and
electron have the same speed of light.  However, in order to obtain a
relativistic continuum limit in general it is necessary to make $\vert
t\vert$ a function of $e^2$.  We shall find that in the limit where
$e^2$ is large, the speed of light is proportional to $\vert t\vert/e$
and it is necessary that $\vert t\vert \sim e$ as $e^2\rightarrow
\infty$.

Equation (\ref{3.3}) reduces to the standard Hamiltonian of QED in the
naive weak coupling continuum limit. In the strong coupling limit
compact QED is confining \cite{poly} and the phase transition
associated with chiral symmetry breaking is generally of first order
\cite{kogut1987}.  In (\ref{3.3}) we have introduced $N_{\tiny L}$
flavors of lattice Fermions labelled by the index $a=1,\ldots,N_{\tiny
L}$.

The Hamiltonian is supplemented with the constraint of gauge
invariance.  The gauge transformations of the dynamical variables,
\bq
\Lambda:&&A_i(x)\rightarrow A_i(x)+\nabla_i\Lambda(x)
\nonumber\\
\Lambda:&&E_i(x)\rightarrow E_i(x)
\nonumber\\
\Lambda:&&\psi^a(x)\rightarrow e^{i\Lambda(x)}\psi^a(x)
\nonumber\\
\Lambda:&&\psi^{a\dagger}(x)\rightarrow\psi^{a\dagger}(x)e^{-i\Lambda(x)}
\label{3.5}
\eq
are generated by the operator
\bq
{\cal G}_\Lambda\equiv\sum_x
\left(-\nabla_i\Lambda(x) E_i(x)+\Lambda(x)
\left(\psi^{a\dagger}(x)\psi^a(x)-N_{\tiny L}/2\right)\right)
\label{3.6}
\eq
The local generator of gauge transformations where $\Lambda$ has
compact support is
\bq
{\partial{\cal G}_\Lambda\over\partial\Lambda(x)}\equiv{\cal
G}(x)=\hat\nabla\cdot E(x)+\psi^{a\dagger}(x)\psi^a(x)-N_{\tiny L}/2
\label{gauss}
\eq
Both (\ref{3.6}) and (\ref{gauss}) commute with the Hamiltonians in
(\ref{3.3}).

In (\ref{3.6}) and (\ref{gauss}) we have subtracted the constant
$N_{\tiny L}/2$ from the charge density operator in order to make the
gauge generator odd under the usual charge conjugation transformation
\bq
\xi:&&A_i(x)~\rightarrow ~-A_i(x)
\nonumber\\
\xi:&&E_i(x)~\rightarrow ~-E_i(x)
\nonumber\\
\xi:&&\psi^a(x)~\rightarrow ~(-1)^{\sum_{k=1}^{D}x_k}\psi^{a\dagger}(x)
\nonumber\\
\xi:&&\psi^{a\dagger}(x)~\rightarrow~ (-1)^{\sum_{k=1}^{D}x_k}\psi^a(x)
\label{xi}
\eq
In fact, the Fermionic charge term in (\ref{3.6}) can be put in the
manifestly odd form ${1\over2}[\psi^{a\dagger}(x),\psi^a(x)]$. Of
course, charge conjugation symmetry of the lattice theory is necessary
to obtain charge conjugation of the continuum theory.  We shall see
later that, particularly at strong coupling, the subtraction term in
(\ref{3.6}) plays an important role.  It seems to have been ignored in
previous literature (for example, see
\cite{kogut1,Kogutreview}).  Its presence is particularly important
when $N_{\tiny L}$ is odd since the charge operator has no zero
eigenvalues in that case (the eigenvalues of
$\psi^{a\dagger}(x)\psi^a(x)$ are integers).

Chiral symmetry is related to translation invariance by one site.  The
Hamiltonian (\ref{3.3}) is invariant under the
transformations
\bq
\chi_j:&&A_i(x)~\rightarrow~ A_i(x+\hat j)
\nonumber\\
\chi_j:&&E_i(x)~\rightarrow~ E_i(x+\hat j)
\nonumber\\
\chi_j:&&\psi^a(x)~\rightarrow ~(-1)^{\sum_{k=j+1}^{D}x_k}
\psi^a(x+\hat j)
\nonumber\\
\chi_j:&&\psi^{a\dagger}(x)~\rightarrow ~(-1)^{\sum_{k=j+1}^{D}x_k}
\psi^{a\dagger}(x+\hat j)
\label{chi}
\eq
for $j=1,\cdots, D$.

In the following we shall use the charge conjugation symmetry which is
a combination of these two transformations:
\bq
C\equiv\xi\chi_1
\label{C}
\eq
This is necessary if the mass operators which we defined in the previous
section are to be
invariant under charge conjugation symmetry.  Also, we shall see that
the strong coupling ground state is invariant under $C$ but not under
either $\xi$ or $\chi_j$ by themselves.

The dynamical problem of Hamiltonian lattice gauge theory is to find
the eigenstates of the Hamiltonian operator (\ref{3.3})
and out of those eigenstates to find the ones which are
gauge invariant, i.e. which obey the physical state condition (or, the
``Gauss' law'' constraint)
\bq
{\cal G}(x)\vert \Psi_{\rm phys.}>=0
\label{3.7}
\eq
In the case of compact
QED, the periodicity of the gauge potential induces an additional symmetry
which, being a large
gauge symmetry, can be represented projectively. In the Schr\"odinger picture
we shall require that the
quantum states which are functions of a configuration of the gauge
field transform as
\bq
\vert A_i(x)+2\pi n>=\exp\left(in\theta[x,i]\right)\vert A_i(x)>
\label{THETA}
\eq
There is a separate parameter $\theta[x,i]$ for each link of the
lattice.  These parameters originate in a way similar to the
theta--angle in ordinary QCD.  The symmetry (\ref{THETA}) together
with the commutator (\ref{3.2}) imply that the spectrum of the
electric field operator has eigenvalues which are separated by
integers and offset by $\theta$:
\bq
{\rm spectrum}[E_i(x)]=\{\theta[x,i]+{\rm integers}\}
\label{specE}
\eq

The Hamiltonian and gauge constraints can be obtained from the gauge
invariant Lagrangian
\bq
L=\sum_x\psi^{a\dagger}(x)(i\partial_t-A_0(x))\psi^a(x)+\sum_{[x,i]}
E_i(x)\dot A_i(x)\nonumber\\ +\sum_{[x,i]} E_i(x)\nabla_i
A_0(x)+\sum_xA_0(x)N_{\tiny L}/2-H
\label{3.8}
\eq
where the temporal component of the gauge field has been introduced to
enforce the gauge constraint and the time derivative terms give the
correct symplectic structure. In order to get Lorentz
invariance of the Fermion spectrum in the weak coupling (naive) continuum
limit, we require half-filling of the Fermionic states, i.e. that the total
charge defined by
$$
\sum_x \left( \psi^{a\dagger}(x)\psi^a(x)-N_{\tiny L}/2\right)
$$
has zero vacuum expectation value.

Here we have considered massless QED.  As well as the gauge invariance
and charge conjugation invariance discussed above, the Hamiltonian is
symmetric under the discrete chiral transformations (\ref{chi}) which
on the lattice corresponds to symmetry under translation by one site.
In later Sections, we shall consider the possibility of spontaneous
breaking of this symmetry.

\subsection{\bf Gauge fixing and quantization}

We shall quantize the gauge fields in the Schr\"odinger picture.  The
quantum states are functions of the link operators which are
represented by the classical variables $A_i(x)$ and the electric field
operators are derivatives
\bq
E_i(x)\equiv -i{\partial\over\partial A_i(x)}
\label{3.9}
\eq
We shall also consider the usual Fock representation of the Fermion
anticommutator.  The empty vacuum is the cyclic vector
\bq
\psi^a(x)\vert 0>=0~~~\forall a,x
\label{3.10}
\eq
and Fermions occupying lattice sites are created by $\psi^{a\dagger}(x)$.

In compact QED the spectrum of the gauge generator is discrete and a
state which obeys the physical state condition can be normalized, thus
implying that there is no need for additional gauge conditions.  The
basis wave-functions for compact QED (in the basis where the Fermions
density and eletric field operators are diagonal)
are $\Psi[n(x)]\Phi[A]$ with the Fermion states
\bq
\Psi[n(x)] = \prod_x \prod_{a=1}^{N_{\tiny L}}
(\psi^{a\dagger}(x))^{n_a(x)} \vert 0>
\label{fermionstates}
\eq
labeled by vectors $n(x)=(n_1(x),\cdots,n_{N_{\tiny L}}(x))$ with
$n_a(x)=0$ or $1$, and the photon states
\bq
\Phi[A]=\exp\left(i\sum e_i(x)A_i(x)\right)
\label{states}
\eq
where the eigenvalues $e_i(x)$ of the electric field operator are in
spectrum$[E_i(x)]$ (\ref{specE}).  Furthermore the states of the
photon field are normalized using the inner product
\bq
<\Phi_1[A],\Phi_2[A]>=
\prod_{[x,i]}\int_0^{2\pi}{dA_i(x)\over2\pi}\Phi_1^{\dagger}[A]\Phi_2[A]
\eq
and the Fermion states have conventional inner product given by
$<0\vert 0>=1$ and the canonical anticommutator relations of the
fermion field operators. The physical state condition (\ref{3.7})
gives the additional restriction that
\bq
\hat\nabla_ie_i(x)=-\rho(x)= -\sum_x\sum_a(n_a(x)-1/2)
\eq
where $\rho(x)$ is the charge density (i.e. the eigenvalue of
$\psi^{a\dagger}(x)\psi^a(x)-N_{\tiny L}/2$).  Pictorially, we can
think of this as containing lines of electric flux joining sites whose
charges are non-zero and also closed loops of electric flux. In the
strong coupling limit, the Hamiltonian is diagonal in the basis
(\ref{states}). This gives a natural explanation of confinement in
compact QED in the strong coupling region.  If we add a
particle--antiparticle pair to a state in (\ref{states}) it must be
accompanied by at least a single line of electric flux.  The energy of
such a line of flux is proportional to its length.  Therefore the
electron-positron interaction grows linearly with distance and is
confining.  This is in contrast to the situation in non-compact QED
where the electric flux is not quantized.  In that case, a
particle--antiparticle pair can have many lines with arbitrarily small
flux. The energy of the field is minimized by the usual Coulomb dipole
configuration.  In high enough dimensions this is not a confining
interaction.

\section{\bf Strong Coupling Expansion}

In compact QED the eigenstates of the electric field operator are
normalizable and can be used for the ground state.
In this case we separate the Hamiltonian into three terms,
\bq
H_0=\sum_{[x,i]}\frac{e^2}{2}E_i^2(x)\nonumber\\
H_1=\sum_{[x,i]}t_{[x,i]}\left(\psi^{\dagger}(x+\hat
i)e^{iA_i(x)}\psi(x)+{\rm h.c.}\right)
\nonumber\\
H_2=\sum_{[x,i,j]}\frac{2}{e^2}\sin^2\left(B[x,i,j]/2\right)
\eq
In the strong coupling limit it is necessary to solve Gauss' law
(\ref{3.7}) for the electric fields and the charge distribution in
such a way as to minimize $H_0$.

When $N_{\tiny L}$ is even, the charge operator has zero eigenvalues
and Gauss' law has the solution where $E_i(x)=0$, which is an obvious
minimum of the $H_0$, and there are $N_{\tiny L}/2$ fermions on each
site. The degeneracy of this state may be
resolved , resulting in an effective Hamiltonian
 which describes the SU($N_{\tiny L}$) antiferromagnet in
the representation with Young Tableau having one column with $N_{\tiny
L}/2$ boxes.
This system has N\'eel order in $D\geq2$ if
$N_{\tiny L}$ is small enough and the chiral symmetry of
electrodynamics is broken, along with the SU($N_{\tiny L}$) flavor
symmetry \cite {dsls}.

In order to see this we follow \cite{dsls} and
resolve the degeneracy by diagonalizing
perturbations in the hopping parameter expansion.  The first order
perturbations vanish.  The first non--trivial order is second order,
\bq
\delta_2= -<g.s.\vert H_1{1\over H_0-E_0}H_1\vert g.s.>
\eq
This matrix element can be evaluated by noting that $H_1$ creates an
eigenstate of $H_0$ different from the ground states with additional
energy
\bq
\Delta E=\frac{e^2}{2}+\frac{e^2}{2}(D)\nabla_1
(x\vert\frac{1}{-\nabla\cdot\hat\nabla}\vert x)~=~e^2
\eq
Diagonalizing the matrix of second order perturbations is equivalent
to finding the spectrum of the effective Hamiltonian
\bq
H_{\rm eff}={2\vert t\vert^2\over e^2}\sum_{[x,i]}\psi^{b\dagger}(x+\hat
i)\psi^b(x)\psi^{a\dagger}(x)\psi^a(x+\hat i)
\nonumber\\
=\frac{2\vert t\vert^2}{e^2}\sum_{[x,i]}J_{ab}(x)J_{ba}(x+\hat i)
\label{heff}
\eq
where the operators
$J_{ab}(x)=\psi^{a\dagger}(x)\psi^b(x)-\frac{1}{2}\delta_{ab}$, are
the generators of the U($N_{\tiny L}$) given in equation
(\ref{generators}) and obeying the Lie algebra in equation
(\ref{algebra}).

The constraint on the total occupation number of each site, $$
\rho(x)=\sum_{a=1}^{N_L} J_{aa}(x) \sim 0
$$ reduces to SU($N_{\tiny L}$) and projects onto a fundamental
representations of SU($N_{\tiny L}$) \cite {dsls}.

It is straightforward to see that the higher orders in the hopping
parameter expansion also have higher orders of $1/e^2$.  In fact, if
we consider the following limit,
\bq
e^2\rightarrow\infty~~~,~~\vert t\vert^2\rightarrow\infty
\nonumber\\
\vert t\vert^2/e^2=~{\rm constant}
\label{limit}
\eq
all higher order perturbative contributions to both the wavefunction
and the energy vanish.  Thus, in this limit, QED is {\bf exactly}
equivalent to an SU($N_{\tiny L}$) antiferromagnet.  That
(\ref{limit}) is the correct limit to take can be seen from the fact
that, if the antiferromagnet in (\ref{heff}) is in an ordered state,
the speed of the spin-waves, which are the gapless low-energy
excitations is proportional to $\vert t\vert/e$.  They have linear
dispersion relation $\omega(k)\sim\vert k\vert$ and play the role of
massless goldstone bosons for broken flavor symmetry.  Their speed
should be equal to the speed of light, which is one in our units.
This implies that $\vert t\vert/e$ should be adjusted so that the
spin-wave spectrum is relativistic, $\omega(k)=\vert k\vert$.  Hence
the limit in (\ref{limit}).

When $N_{\tiny L}=2$, this model is the quantum Heisenberg
antiferromagnet in the $j=1/2$ representation.  It is known to have a
N\'eel ordered ground state in $D\geq3$ \cite{lieb} and there is good
numerical evidence that it has N\'eel order in $D=2$.  The
antiferromagnetic order parameter is the mass operator
\bq
\vec\Sigma = \sum_x (-1)^{\sum_{i=1}^D x_i}\psi^{\dagger}(x)\vec\sigma\psi(x)
\eq
which obtains a vacuum expectation value in the infinite volume limit.
Thus, when $N_L=2$ the strong coupling limit breaks chiral symmetry
and generates electron mass.  It is interesting that in this case
there is an iso--vector condensate.  In the strong coupling limit this
seems unavoidable.  The only way to get an iso--scalar condensate is
with a charge density wave.  However such a configuration always has
infinite coulomb energy compared to an electric charge neutral but
isospin carrying condensate.

The low energy excitations of this systems (with energies of order
$\vert t\vert^2/e^2$ are spin waves.  All other excitations have
energies which go to infinity in the limit (\ref{limit}).  The spin
waves are the pions which are the scalar Goldstone bosons arising from
spontaneous breaking of the vector flavor symmetry
SU(2)$\rightarrow$U(1).

For large $N_{\tiny L}$ there is some evidence that the SU$(N_{\tiny
L})$ antiferromagnet in these specific representations has a
disordered ground state \cite{read2}.  Particularly in 2+1-dimensions
it is known that for infinite $N_L$ the ground state is disordered
\cite{Affleck 1988}.

When $N_{\tiny L}$ is odd, since the charge density operator has no
non-zero eigenvalues, it is impossible to find a zero eigenstate of
the Gauss' law constraint operator without some electric field.  The
problem which we must solve is to minimize the energy functional $
\sum E^2$ subject to the constraint $\hat\nabla\cdot E=-\rho$ where at
each site $\rho$ has one of the values $$ -\frac{N_{\tiny
L}}{2},-\frac{N_{\tiny L}}{2}+1,\ldots,\frac{N_{\tiny
L}}{2}~~~N_{\tiny L}~{\rm an~odd~integer} $$ The reader can easily
convince herself that the charge distribution and electric field which
one obtains are \cite {dsls}, $$
\rho_0(x)={1\over2}(-1)^{\sum_{i=1}^D x_i}~~~E_i(x)=
\frac{1}{4D}(-1)^{\sum_{i=1}^D x_i}$$
 These configurations break chiral symmetry in that they are not
invariant under the transformation $\chi_1$ in (\ref{chi}) but they
are symmetric under charge conjugation $C$ defined in (\ref{C}).

The ground state energy per lattice site is
\bq
\frac{E_{0(\rm coul)}}{|V|}=\frac{e^2}{32D}
\eq
Note that it is of order $e^2$.  This is in contrast to the ground
state energy when the number of lattice Fermion flavors is even, which
is of order $\vert t\vert^2/e^2\sim 1$.

The ground states that we have found are highly degenerate in that
only the number of Fermions at each site is fixed.  Their quantum
state can still take up any orientation in the vector space which
carries the representation of the flavor SU($N_{\tiny L}$) given by
the Young Tableaux with one column and $(N_{\tiny L}-1)/2$ or
$(N_{\tiny L}+1)/2$ rows on each of the two sublattices $A$ and $B$ in
which we have divided the lattice. Namely : $A$ is the set of all
points where $\sum_i x_i$ is even and $B$ the one where $\sum_i x_i$
is odd. The differing occupation numbers on sites on each sublattice
yield different representations of SU$(N_L)$.

The degeneracy must be resolved by diagonalizing the perturbations,
which are non-zero in second order and the problem is equivalent to
diagonalizing the antiferromagnet Hamiltonian (\ref{heff}).  Also, in
the limit (\ref{limit}) this correspondence is {\bf exact}.

The strong coupling ground states that we find when $N_{\tiny L}$ is
odd contains a charge density wave.  The staggered charge density
operator has expectation values
\bq
\frac{1}{|V|}<\sum_x(-1)^{\sum_{k=1}^{D}x_k}\psi^{\dagger}(x)\psi(x)>=
\pm \frac{1}{2}
\eq
This condensate is an isoscalar and we have shown that it must always
occur in all dimensions.  When $N_{\tiny L}>1$ mass operators with
certain generators of SU($N_{\tiny L}$) could have expectation values
if the ground state has antiferromagnetic order.  However, unlike the
case of even $N_{\tiny L}$, the antiferromagnetic order is not
required in order to have chiral symmetry breaking.

The ground state we find breaks chiral symmetry.  This is a true
dynamical symmetry breaking since, in infinite volume, the ground
states which are related by a chiral transformation are never mixed in
any order of strong coupling perturbation theory.  Furthermore, there
are no local operators which couple them.

We conclude that the strong coupling ground state breaks chiral
symmetry for any odd $N_{\tiny L}$ and in any dimensions.  As in the
case of even $N_L$ there is also the possibility (and for small $N_L$
the likelyhood) that the SU$(N_L)$ antiferromagnet we obtain here is
in a N\'eel state and the flavor symmetry is also broken \cite{read2}.

Notice that in the ground state, the electric fields are not integers,
but on each link, the spectrum of the electric field operator is
$1/4D$+ integers. The ``theta angles'' 1/4D survive all orders in
strong coupling perturbation theory.

\section{\bf Remarks}

In this paper we reviewed our approach \cite {dsls} to analyze
the strong coupling limit of quantum electrodynamics using
the Hamiltonian picture and a lattice regularization.  We showed also the
mechanism for chiral symmetry breaking in this lattice gauge theory. We
chose to use staggered Fermions because they give the closest
analog to interesting condensed matter physics systems.

In 1+1 dimensions, staggered Fermions give $N_{\tiny L}$ species of 2-
component Dirac Fermions.  When $N_{\tiny L}=1$ we obtain the
Schwinger model with a lattice regularization.  Also, in this case, we
have found that the chiral symmetry is broken dynamically.  Of course,
due to the staggered Fermion regularization there is no continuous
chiral symmetry, which is as it should be since it should be
impossible to regularize the Schwinger model so that there is
simultaneous continuous chiral and gauge symmetry.  However, to match
the solution of the continuum Schwinger model, the Fermion should
obtain mass.  This indeed happens in our strong coupling limit by
spontaneous symmetry breaking.  (Although we disagree with some
aspects of the formalism, we agree with the results of reference
\cite{kogut1} on this point.)

This result should not be confined to strong coupling, but should
persist for all coupling, i.e. the critical coupling in $D=1$ should
be at $e^2=0$.  We conjecture that this sort of symmetry breaking for
small $e^2$ is a manifestation of the Peierls instability --- the
tendency of a one dimensional Fermi gas to form a gap at the Fermi
surface.  This happens with any infinitesimal interaction.

In fact, this must also happen for the case where $N_{\tiny L}$ is
even.  Then, there cannot be any spin order in 1 dimension.  However,
anomalies break the isoscalar chiral symmetry in the continuum theory
and should also do so here.  This means that there should be a
dynamical generation of charge density wave which would be driven by
the Peierls insability.  It also implies that for $N_{\tiny L}=2$ for
example, the ground state in the strong coupling limit would not be a
Heisenberg antiferromagnet, but would be alternating empty site and
site with two electrons in a spin singlet state.  This state, even
though it has large coulomb energy, avoids the infrared divergences of
gapless Fermions.

In higher dimensions, $D\geq2$, one should explore the possibility of
phase transitions between different symmetry breaking patterns for the
SU($N_{\tiny L}$) flavor symmetry as one varies $N_{\tiny L}$.  There
is already some work on this subject in the condensed matter physics
literature on SU($N_{\tiny L}$) antiferromagnets \cite{read2}.  They
analyze the SU($N_{\tiny L}$) antiferromagnet which is similar to the
strong coupling limit of an U($N_{\tiny C}$) gauge theory (see
\cite{dsls} for details) which is in the representation corresponding
to a rectangular Young Tableau with $N_{\tiny L}$ rows and $N_{\tiny
C}$ columns.  They work in the large $N_{\tiny C}$ limit and show that
there is a phase transition from the spin ordered N\'eel phase to a
disordered phase when $N_{\tiny L}\sim N_{\tiny C}$.  For QED
$N_{\tiny C}=1$ so their analysis is not accurate.  For odd $N_{\tiny
L}$ the chiral symmetry is always broken and the question we are
asking is whether the flavor symmetry is also broken.  For even
$N_{\tiny L}$ possible phase transition is relevant to both chiral and
flavor symmetry breaking. In this context it is interesting to
ascertain if the order--disorder phase transition which occurs as one
increases $N_{\tiny L}$ in the SU($N_{\tiny L}$) antiferromagnet is
the same one that appears in the study of chiral symmetry breaking in
2+1--dimensional QED in the continuum
\cite{Applequist,Nash} where they find that chiral symmetry is broken
only if the number of flavors is less than a certain critical value.
These questions have been addressed in \cite {diasod}.

Our study suggests that, besides the critical $N_{\tiny L}$, for a
fixed $N_{\tiny L}$ which is small enough, there should also be a
critical coupling constant $e^2$ and, in fact, a critical line in the
$N_{\tiny L}$--$e^2$ plane where there is a second order phase
transition between a spin ordered chiral symmetry breaking phase and a
disordered (and possibly chirally symmetric phase).  We speculate that
in 3+1--dimensions a similar situation could occur.  The latter
phenomenon has recently been discussed in the context of
2+1-dimensional gauge theory \cite{diasod}.  It can in principle shed
light on the nature of the phase transition in the lattice theories.
It would be interesting to analyze the critical behavior further,
particularly in the lattice theory, since one could address the
question of whether this phase transition is the same as the one seen
in Josephson junction arrays.

We have seen that the Coulomb gas representation of classical
Josephson junction arrays is described at infinite coupling by the
$N_L=1$ Coulomb gas. Away from infinite coupling in the case of
electrodynamics and away from the classical limit in the case
of junctions arrays the two systems may still have many common
features.

In the quantum regime, Josephson junction arrays may be
modeled by the quantum frustrated XY
model which is equivalent to a model
of lattice electrons in a magnetic field determined
by the charge density \cite{diaso}.  In a mean field theory, where
the charge
density is replaced by its average, this resembles the
Azbel-Wannier-Hofstaeder problem for lattice electrons with nearest
neighbor hopping with an external magnetic field \cite{diaso}. When
$f=1/2$ the low energy components of the electron
spectrum are approximately described by a massless Dirac Hamiltonian.
Thus, for $f=1/2$, the model has relativistic lattice electrons and is
gauge and parity invariant. When $f$ is a rational number,
the frustrated
quantum XY model admits a flux phase as its mean field theory ground
state \cite{diaso}. It would be interesting to analyze the properties of this
model beyond mean field theory. In particular to ascertain
if - for $f$ rational - the flux phase may be regarded as the exact
ground state \cite{marston}
as well as to investigate if, for $f=1/2$, quantum fluctuations
could reproduce the relevant operators of compact QED.


\newpage
\begin{thebibliography}{99}
\bibitem{s}G. W. Semenoff, Mod. Phys. Lett. A17, (1992) 2811;
E. Langmann and G. W. Semenoff, Phys. Lett. B297 (1992), 175.
\bibitem {dsls} M.C. Diamantini, E. Langmann, G.W. Semenoff, P. Sodano," SU(N)
Antiferromagnets and the Phase Structure of QED in the Strong Coupling
Limit" , DFUPG 69-92,December 1992;\\ M. C. Diamantini, P. Sodano, E.
Langmann and G. W. Semenoff, ``Quantum Spin Systems and the Vacuum
Structure of Strong Coupling Gauge Theory with Dynamical Fermions'',
Proceedings of {\it Field Theory and Collective Phenomena}, Perugia,
Italy, June 1992.
\bibitem{smit}J. Smit, Nucl. Phys. B175, (1980) 307.
\bibitem{fradbk}E. Fradkin, {\it Field Theories and Condensed Matter
Systems}, Frontiers in Physics, Addison-Wesley Publishing Company,
1991.
\bibitem{Affleck 1988}I. Affleck and B. Marston, Phys. Rev. B37, (1988) 3774.
\bibitem {azha} I. Affleck, Z. Zou, T. Hsu, and P.W. Anderson,Phys. Rev.B38,
(1988) 745.
\bibitem {hop} D. Hofstaeder, Phys.Rev. B14, (1976) 2239.
\bibitem {ralu} R. Rammel, T.C. Lubensky and G. Toulouse,
Phys.Rev. B27, (1983) 282 0.
\bibitem {shst} W. Shih and D. Stroud, Phys. Rev. B28, (1983) 6575.
\bibitem{halsey}T. C. Halsey, Phys. Rev. B31, (1985) 5728;\\
T. C. Halsey, J. Phys. C 18, (1985) 2437.
\bibitem {scbi} D. Schmeltzer, A.R. Bishop, Phys. Rev.B41, (1990) 9603.
\bibitem {diaso} M.C. Diamantini, P. Sodano, Phys.Rev.B45, (1992) 5737.
\bibitem{jknn}J. Villain, J. Phys. 36 (1975), 581; J.V. Jose,
L. P. Kadanoff, S. Kirkpatrick and D. R. Nelson, Phys. Rev. B16
(1977), 121.
\bibitem{bishop}A. R. Bishop and P. S. Lomdahl, Phys. Rev. B41 (1990), 10983.
\bibitem{xxx}C. Bruder, R. Fazio, A. Kampf, A. van Otterlo
and G. Sch\"on, `Quantum phase transitions and commensurability in
frustrated Josephson junction arrays'', Karlsrue preprint, 1992.
\bibitem{fomin et.al.}P. Fomin, V. Gusynin, V. Miranski and Yu. Sitenko, Riv.
Nuov. Cim.  6, (1983) 1.
\bibitem{miransky}V. Miransky, Nuov. Cim. 90A, (1985) 149.
\bibitem {diasod} M.C. Diamantini, G.W. Semenoff, P. Sodano ," Chiral Symmetry
Breaking in Three Dimensional Gauge Theories" , DFUPG 70-92,December
1992
\bibitem{Applequist}Applequist, D. Nash and L. C. R. Wijewardhana,
Phys. Rev. Lett. 60, (1988) 2575.
\bibitem{Nash}D. Nash, Phys. Rev. Lett. 62, (1989) 3024.
\bibitem{kogutnew}J. Kogut, Proceeding of Lattice'92, Amsterdam, June,
1992; E. Dagotto, J. Kogut and A. Kocic, Phys. Rev. Lett. 62, (1992)
1083.
\bibitem{poly}A. Polyakov, Nucl. Phys. B120, (1977) 429.
\bibitem{niemi}A. J. Niemi and G. W. Semenoff, Phys. Rev. Lett. 51,
(1983) 2077; A. N. Redlich, Phys. Rev. Lett. 52, (1984) 1.
\bibitem{Kluberg}H. Kluberg-Stern, A. Morel, O. Napoly and B. Petersson,
Nucl. Phys.  B220 [FS8], (1983) 447.
\bibitem{Kogutreview}J. B. Kogut, Rev. Mod. Phys.  55, (1983) 775.
\bibitem{kogut1987}J. B. Kogut and E. Dagotto, Phys. Rev. Lett. 59, (1987)
617.
\bibitem{kogut1}A. Caroll, J. Kogut, D. K. Sinclair and L. Susskind, Phys.
Rev. D8, (1976) 2270;\\T. Banks, L. Susskind and J. Kogut, Phys. Rev.
D4, (1976) 1043.
\bibitem{lieb}T. Kennedy, E. H. Lieb and B. S. Shastry, J. Stat.
Phys. 53, (1988) 1019.
\bibitem{read2}N. Read and S. Sachdev, Phys. Rev. Lett. 62 (1989) 1694;\\
N. Read and S. Sachdev, Phys. Rev. B42 (1990) 4568;\\ N. Read and S.
Sachdev, Nucl. Phys. B316 (1990) 609.
\bibitem{marston}I. Affleck, D.P. Arovas, J.B. Marston, D.A. Rabson
Nucl.Phys. B366 (1991) 467.


\end{thebibliography}
\end{document}